# Electrically Switchable Metadevices via Graphene

Osman Balci[1*], Nurbek Kakenov[1], Ertugrul Karademir[2], Sinan Balci[3], Semih Cakmakyapan[4],

Emre O. Polat[5], Humeyra Caglayan[6], Ekmel Özbay[1,7], Coskun Kocabas[1,8*]

[1]*Department of Physics, Bilkent University, 06800 Ankara, Turkey*

[2]*School of Physics and CRANN, Trinity College Dublin, College Green, Dublin 2, Ireland*

[3]*Department of Astronautical Engineering, University of Turkish Aeronautical Association, 06790 Ankara, Turkey*

[4]*Electrical Engineering Department, University of California Los Angeles, Los Angeles, California 90095, USA*

[5]*ICFO-Institut de Ciencies Fotoniques, The Barcelona Institute of Science and Technology, 08860 Castelldefels (Barcelona), Spain*

[6]*Department of Electrical and Electronics Engineering, Abdullah Gul University, 38039 Kayseri, Turkey*

[7]*Nanotechnology Research Center—NANOTAM, Bilkent University, 06800 Ankara, Turkey*

[8]*Current address: School of Materials, University of Manchester, M13 9PL Manchester, UK*

*Corresponding authors' email: obalci@fen.bilkent.edu.tr, ckocabas@fen.bilkent.edu.tr*

**Metamaterials bring sub-wavelength resonating structures together to overcome the limitations of conventional materials. The realization of active metadevices has been an outstanding challenge that requires electrically reconfigurable components operating over a broad spectrum with a wide dynamic range. The existing capability of metamaterials, however, is not sufficient to realize this goal. Here, by integrating passive metamaterials with active graphene devices, we demonstrate a new class of electrically controlled active metadevices working in microwave frequencies. The fabricated active metadevices enable efficient control of both amplitude (> 50 dB) and phase (> 90°) of electromagnetic waves. In this hybrid system, graphene operates as a tunable Drude metal that controls the radiation of the passive metamaterials. Furthermore, by integrating individually addressable arrays of metadevices, we demonstrate a new class of spatially varying digital metasurfaces where the local dielectric constant can be reconfigured with applied bias voltages. Additionally, we**

**reconfigure resonance frequency of split ring resonators without changing its amplitude by damping one of the two coupled metasurfaces via graphene. Our approach is general enough to implement various metamaterial systems that could yield new applications ranging from electrically switchable cloaking devices to adaptive camouflage systems.**

## INTRODUCTION

Metamaterials have been a powerful tool to control and manipulate electromagnetic waves and their interaction with matter (*1-3*). The integration of passive metamaterials with a variety of tuning mechanisms has been extensively examined to generate active metadevices that have novel functionalities (*4-7*). Owing to the different tuning mechanisms, these active metadevices can be classified into three general categories, which are circuit-, material-, and physical-based metadevices(*8*). The circuit-based metadevices use variable capacitors or switches to alter the lumped elements of the equivalent circuit. For instance, the reverse bias voltage applied on a varactor integrated on a split ring resonator changes the effective capacitance of the equivalent circuit (*9-12*). Material-based metadevices rely on controlling bulk material properties, such as the permittivity, permeability, or conductivity of the individual unit cell under an external stimulus (*7, 13-18*). The conductivity of semiconducting material deposited in the split gap can be tuned by illumination and thereby it alters the resonance frequency of the metamaterial. Similarly, interconnected metamaterials fabricated on a semiconductor surface have been used to control the depletion area that alters the free carrier absorption (*19, 20*). Alternatively, some phase transition materials such as $VO_2$ have been used to alter the resonance behavior of metamaterials (*21, 22*). Changing the temperature of these materials around the phase transition temperatures, one can control their conductivity and hence the resonance behavior of metamaterials made of or integrated with them. On the other hand, physical-metadevices tune their response by changing their shape or relative position of sub-components (such as split gap) (*23-27*). For example, micro electro mechanical (MEMs) based devices can modify the resonance frequency of metadevices in terahertz spectra (*25*). Although the demonstrated devices provide some degree of tunability, their performances are limited to narrow spectra with a small dynamic range due to the material and fabrication limitations. Therefore, these technologies would greatly benefit from a material that yields large tunability over broad spectra. None of the existing materials provide these challenging requirements. Furthermore, the requirement for electrically controlled tunability places another challenge for practical applications of metadevices.

2-dimentional crystals provide new perspectives for reconfigurable smart surfaces that can be used for the realization of electrically tunable metadevices (*28-32*). The thickness of 2D crystals is much shorter than the effective wavelength, therefore, their electromagnetic response solely originates from the charge carriers. Recent studies have shown that tuning the density of high mobility free carriers on graphene yields an unprecedented ability to control light-matter interaction over a very broad spectrum ranging from visible to microwave frequencies (*33, 34*). By engineering the shape or doping level, the plasma frequency of graphene can be tuned between IR and THz frequencies (*30, 35*). Notably, the charge density alters the effective dielectric constant of the medium. In another approach, graphene is coated on metamaterials (*36, 37*) or optical antennas (*38, 39*) to yield electrically tunable metadevices. However, these devices are not suitable for practical applications due to limited tunability (*40*). Recently, we discovered a simple device structure consisting of an electrolyte medium sandwiched between two large area graphene electrodes. This geometry permits an efficient mutual gating between two graphene electrodes that yields charge densities on the order of $10^{14}$ cm$^{-2}$ (*41*). Using this supercapacitor structure, we fabricated various optoelectronic devices including optical modulators (*41, 42*), electrochromic devices (*43*), and switchable radar absorbing surfaces (*44*). In this paper, using graphene supercapacitors incorporated with metallic split ring resonators, we demonstrated a new type of electrically tunable metadevices. The fabricated metadevices are based on gate tunable high mobility free carriers on graphene, which introduces electrically tunable dissipation in the resonator that is capacitively coupled to the graphene electrodes.

**RESULTS**

Integrating split ring resonators (SRR) in close proximity to graphene surface yields a new type of hybrid metamaterial whose resonance amplitude can be tuned by various means. Previous attempts to integrate graphene with metamaterials yielded very limited modulation in infrared and terahertz frequencies (*35, 37*). Here, we follow a different approach using microwave metamaterials capacitively coupled to a large area graphene that yields substantial tunability. Fig. 1A shows a schematic drawing of this hybrid structure. The presence of graphene in close proximity to SRR introduces additional electrical losses due to the sheet resistance of graphene. The equivalent small signal model of the hybrid structure is shown in Fig. 1B. The ring and split gap provide the inductance, $L$, and the capacitance, $C$, respectively. The resistor, $R$, models the dissipation on the metal due to the electrical and radiation resistance. In our design, the SRR is capacitively coupled to graphene ($C_c = \frac{\varepsilon \varepsilon_0}{d}$, the coupling capacitance, $\varepsilon$ is the dielectric constant of the medium and $\varepsilon_0$ is the free space permittivity). The graphene layer can be modeled by the sheet resistance ($R_G = \frac{1}{\sigma(n)}$, where $\sigma(n)$ is a charge dependent sheet conductance of graphene) and quantum capacitance of the graphene electrodes ($C_Q = \frac{2e^2}{\hbar \upsilon_F} \sqrt{\frac{n}{\pi}}$ where $e$ is the elementary charge, $\upsilon_F$ is the Fermi velocity, n is the charge density). The quantum capacitance of unintentionally doped graphene layer ($C_Q \sim 0.5$ μF/cm$^2$) is much larger than the serial coupling capacitance ($C_c \sim 5$ pF/cm$^2$) and, therefore, in the small signal model we can neglect its contribution (*45*). Capacitive coupling of a graphene layer to an SRR alters the required total impedance to achieve a resonance in transmitting electromagnetic waves. Therefore, both the capacitive coupling distance (d$_1$) and the charge density (or bias voltage) on graphene change the resonance condition.

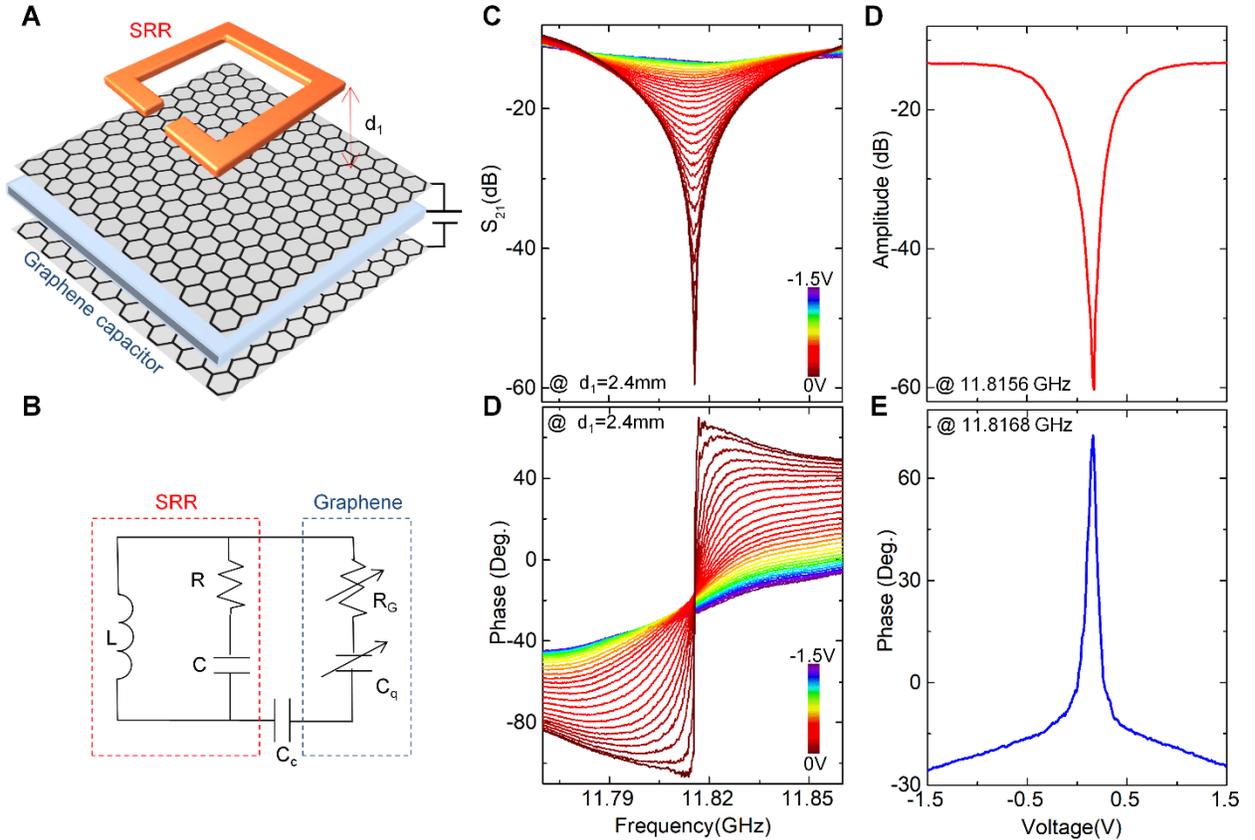

**Fig. 1. Electrically tunable metadevices.** (**A**) Schematic representation of the hybrid metamaterial system consisting of split-ring resonator capacitively coupled to the graphene electrodes. The capacitive coupling is defined by the SRR-graphene separation, $d_1$. (**B**) Small-signal equivalent circuit model of the metadevice. SRR is represented by the $L$, $R$, $C$ lump circuit elements; the graphene layer is modeled by the variable sheet resistance, $R_G$ and quantum capacitance, $C_Q$. $C_c$ models the capacitive coupling. (**C, D**) Spectra of the magnitude and phase of the transmittance ($S_{21}$) at various bias voltages. The color bar shows the bias voltage. (**E, F**) The variation of the amplitude and phase of the transmittance with the bias voltage.

In order to tune the electrical resonance of metamaterials we varied the charge density on graphene layer via ionic gating. It should be emphasized here that the technical challenge for graphene based microwave devices is the requirement of large area devices owing to the centimeter scale wavelength (i.e. $\lambda$= 3 cm at 10 GHz). To overcome this challenge, we synthesized large area graphene by chemical vapor deposition (CVD) on copper foils which enables us to realize the proposed microwave metadevices (*46*). Graphene electrodes were synthesized on copper foils using a CVD system and subsequently we transferred them onto flexible PVC substrates by using hot lamination technique. Afterwards, we etched the laminated copper foils in a 5 mM FeCl$_3$ aqueous solution. Although CVD synthesis of graphene enables to grow large area graphene layers, the quality of graphene may diminish due to the inevitable defects and polymeric residues after the transfer. To fabricate graphene capacitor, we first put a paper tissue with a thickness of 80 µm between two graphene electrodes as a spacer layer and then we soaked ionic liquid (1-Butyl-3methlimidazolium hexafluorophosphate) into the tissue. The fabricated device structure resembles the supercapacitors formed by single layer graphene electrodes. The used electrolyte permits mutual electrolyte gating between the graphene electrodes without using any metallic gate electrodes. Indeed, under an external bias voltage, the electrolyte gets polarized and gates the graphene electrodes. The graphene electrode connected to the negative voltage becomes electron-doped (n-doped), whereas the other one becomes hole-doped (p-doped). At zero voltage, the sheet resistance of undoped graphene is around 2.5 k$\Omega$ and it decreases to 0.7 k$\Omega$ at $\pm3$ V. In fact, the charge density on the graphene electrodes changes between $3\times10^{11}$ to $0.5\times10^{14}$ cm$^{-2}$. The metallic SRRs were fabricated on another PVC film by printing the SRR shapes on previously laminated 10 µm thick copper foil followed by the chemical etching of copper. Detailed explanation of device fabrication and characterization procedures are all presented in section S1. After assembling the device, we measured the scattering parameters using a broadband horn antenna as a source and a monopole antenna as a receiver connected to a two-port network analyzer. To excite the electrical resonance, we polarized the electric field along the split gap of SRR. The active device yields an unprecedented ability to control the intensity and phase of the transmitted electromagnetic waves. Actually, Fig. 1 (C and D) show the amplitude and phase of the transmittance ($S_{21}$) measured through the fabricated device at various bias voltages at $d_1$=2.4 mm, respectively. Here, $d_1$=2.4mm is the critical capacitive coupling distance ($d_c$) between SRR arrays and graphene capacitor resulting a maximum modulation of resonance amplitude in dB scale with bias voltage. At 0 V, the device yields a resonance at 11.8156 GHz with a resonance transmittance of -60 dB. When we applied a bias voltage, electrons and holes accumulate on the graphene electrodes and yield significant damping that diminishes the resonant behavior. At 1.5 V, for example, the resonance transmittance is -12 dB. Fig. 1 (E and F) show the voltage dependence of the amplitude of transmittance at resonance (11.8156 GHz) and the phase (at 11.8168 GHz). The phase of the transmitted signal varies from -30° to 70°. Indeed, we observed a symmetric behavior when we changed the polarity of the bias voltage owing to the ambipolar behavior of the graphene electrodes. Noteworthily, the charge neutrality point is slightly shifted to positive voltages due to the slight chemical doping of graphene by polarized ionic liquid. The shift can be minimized by waiting longer during the voltage scan. More details about the microwave performance of our metadevice is explained in section S2.

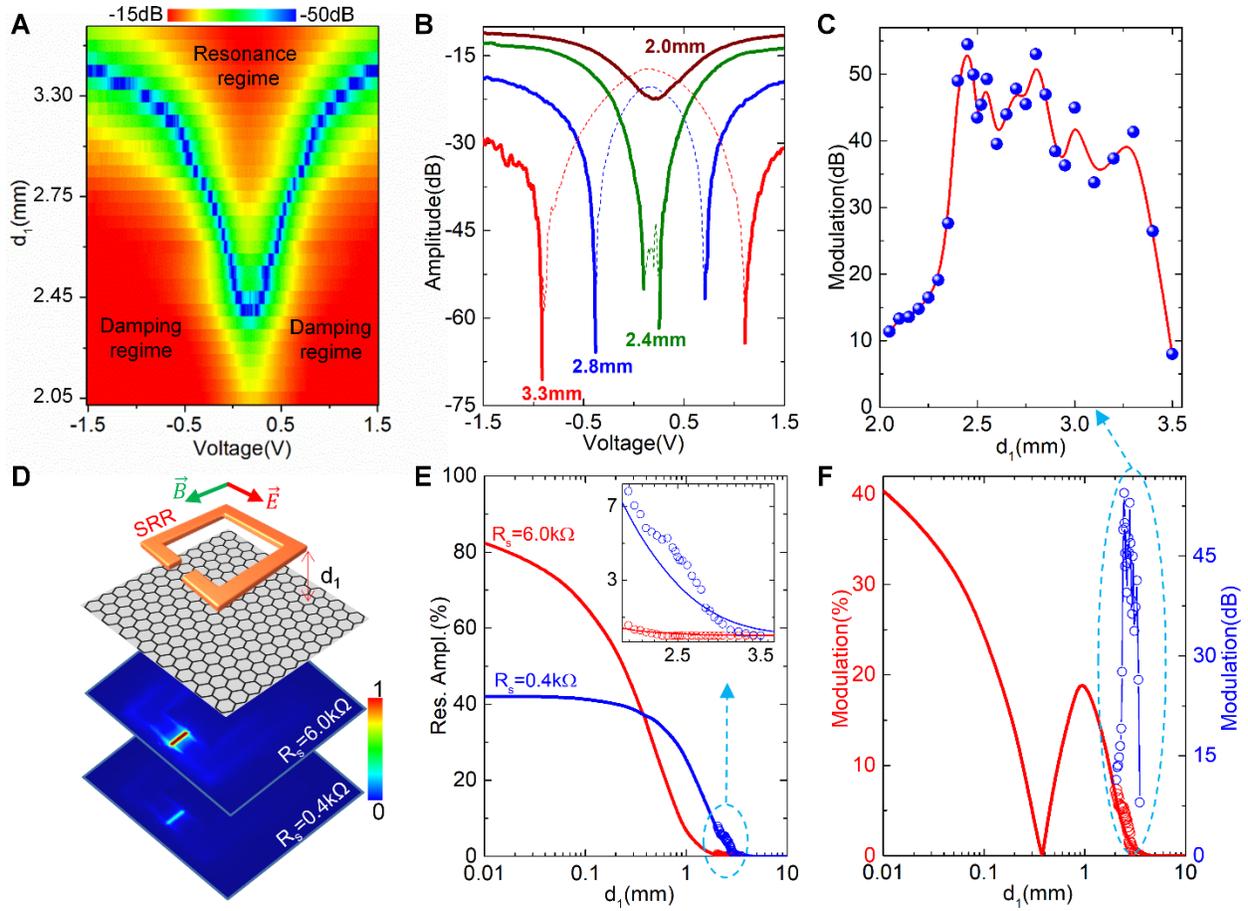

**Fig. 2. Investigating coupling distance dependence ($d_1$) of the metadevice performance.** (**A**) Variation of the resonance amplitude with $d_1$ and bias voltage in the forward direction. While the bias voltage damps the resonance at the damping regimes, it changes the radiation pattern of SRRs without changing its amplitude in the resonance regimes. (**B**) Resonance amplitude as a function of bias voltage at various $d_1$, extracted from a. The solid lines represent damping and dash lines represent resonance regimes. (**C**) Modulation of transmission amplitude at resonance with bias voltage as a function of $d_1$. Modulation is maximum at $d_c$=2.4 mm. (**D**) Schematic drawing for the metadevice structure used in finite element (FE) simulation together with simulated electric field distribution on SRR at 6 kΩ and 0.4 kΩ sheet resistance ($R_s$) of single layer graphene at $d_1$=2 mm. **e,** Simulated variation of resonance amplitude of SRR arrays with $d_1$ for $R_s$=6 kΩ and 0.4 kΩ. The inset shows calculated resonance amplitude variation with experimentally measured ones. **f,** Simulated and measured resonance amplitude modulation as a function of $d_1$. The blue scattered circles are experimentally measured amplitude modulation in dB scale.

To understand the effect of the separation ($d_1$) between graphene capacitor and metamaterial arrays on the performance of our metadevice, we measured near field transmission spectrum in the forward direction at various $d_1$ values while sweeping the bias voltage between ±1.5 V. In fact, Fig. 2A shows the resonance amplitude as a function of $d_1$ and applied bias voltage. We separate the color plot in two regimes; resonance and damping regimes. In resonance regime (above the blue regions), the SRR is not resonating in the forward direction and, therefore, the amplitude increases as $d_1$ goes above $d_c=2.4$ mm. Indeed, the SRR is resonating in the resonance regime but the extinction pattern is not at the position of the monopole antenna. It should be emphasized here that the variation in the near field extinction pattern of SRRs is presented more clearly in Fig. 3. Firstly, the extinction pattern of the SRR is collected towards the antenna position at a critical bias voltage ($V_c$) and subsequently, the damping starts beyond that voltage in the damping regimes. The blue regions in Fig. 2A show $d_1$ and $V_c$ conditions together to achieve a resonance having an amplitude below -50dB in the forward direction at the position of the monopole antenna. In the damping regimes (below the blue region in Fig. 2A), the SRR is always resonating in the forward direction hence applied bias voltage damps its resonance. Fig. 2B shows line plots of the variation in the resonance amplitude with the applied bias voltage for some $d_1$ values, extracted from Fig. 2A. Here the straight lines represent the damping regime and the dash lines represent the resonance regime. Resonance amplitude below -50 dB can be achieved for $d_1 \geq d_c$ and the resonance can be damped with the bias voltages above $V_c$ (varies for each $d_1 \geq d_c$). To find the modulation of the resonance amplitude with the bias voltage, we compare the resonance amplitude at ±1.5 V and $V_c$ for each $d_1$ values. We have achieved the maximum modulation of 55 dB at $d_c=2.4$ mm (Fig. 2C). The modulation increases very fast for $d_1 < d_c$ and decreases slowly for $d_1 > d_c$. More detailed measurements and analysis for the experimental characterization of the coupling distance ($d_1$) are presented in section S2. To further investigate the effect of $d_1$ on the performance of our metadevice, we simulate our device in a finite element simulation platform (COMSOL) using the device structure shown in Fig. 2D. We used a single layer graphene and an SRR coupled to that graphene at $d_1$ in the unit cell. The color plots in Fig. 2D show the normalized electric field intensities on the SRR for 6.0 and 0.4 k$\Omega$ sheet resistances ($R_s$) of the single layer graphene. The field is confined in the split gap and doping single layer graphene decreases the intensity of the electric field. Then, we calculate the resonance amplitude of the SRR as a function of $d_1$ for 6.0 and 0.4 k$\Omega$ $R_s$ values (Fig. 2E). The amplitudes for both $R_s$ values are decreasing with $d_1$ due to the variation in the coupling capacitance. The scattered circles in Fig. 2E and the inset show the measured resonance amplitudes of our metadevice for doped and undoped cases. As a result, there is a good agreement between the simulation and experiments. Using the calculated resonance amplitudes, we calculate the modulation of the amplitude in percent (%) scale as a function of $d_1$ (Fig. 3F). The global maximum is achieved at very short $d_1$ values (let's say at 0 mm) and the local maximum is at $d_1$=1 mm. Our measured modulation in % scale follows the calculated one very well for $d_1$ values used in the experiments. Using the same strategy used for simulation of the metadevices working in the microwave frequencies, we simulated the performance of graphene to control the resonance amplitude of SRRs from terahertz to visible frequencies. In fact, we have found the critical coupling distances to achieve maximum modulation in their amplitude both in dB and % scales. Detailed results of the simulations performed from microwave to visible frequencies are presented in section S3. As a result of the simulations performed, we conclude that our approach is valid from microwave to visible frequencies in the electromagnetic spectrum.

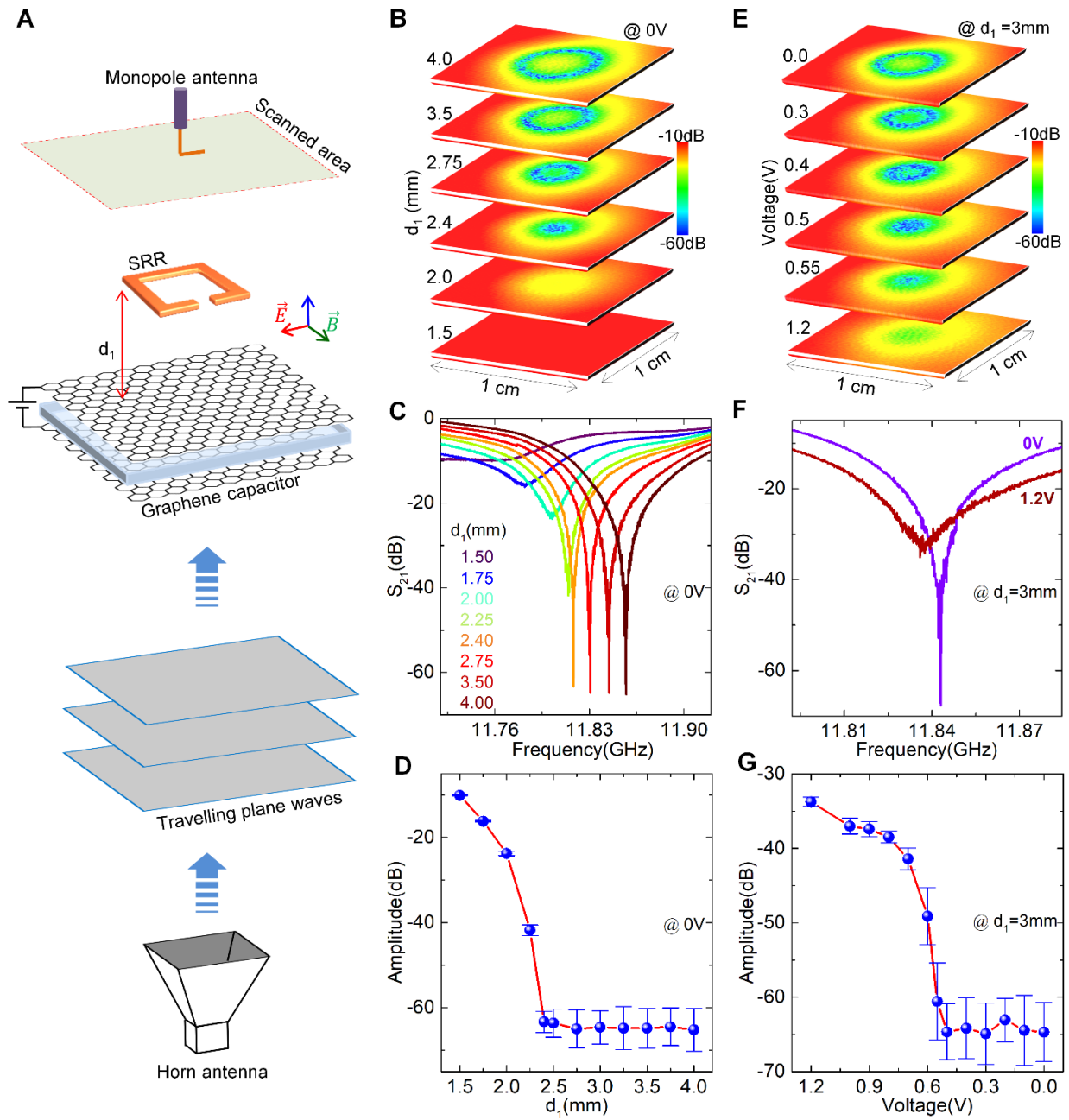

**Fig. 3. Near-field characterization of the active metadevices. (A)** Near-field scanning technique to map the local scattering parameter ($S_{21}$) of the device. Scattering parameter provides the magnitude of the local electric field along the direction of the antenna. **(B to D)** Spatial map, spectrum and average resonance amplitude of the local scattering parameter of the metadevices probed by a monopole antenna at different graphene-SRR distances at 0 V. **(E to G)** Voltage dependence of the spatial map, spectrum, and resonance amplitude of the local scattering parameters of device at $d_1$=3 mm.

To explore the tunability of electric field distribution of active metadevices, we used a near-field scanning technique to map the local electric field distributions (*47*). We placed the device on a computer controlled motorized stage and recorded the spectrum of local scattering parameter by using a monopole antenna connected to a network analyzer, Fig. 3A. First, we examined the critical coupling condition between the graphene electrode and SRR at zero bias voltage. Fig. 3B shows the measured local transmission ($S_{21}$) intensity distribution of metadevices at different graphene-SRR separation. In the near-field of the metadevices, the complete extinction of the microwave signal (i.e. resonance) occurs on a ring-shape region whose radius depends on the graphene-SRR separation. When the graphene is very close (< 2 mm) to the SRR, the resonance diminishes due to the large dissipation on graphene. Fig. 3C shows the transmittance spectrum and Fig. 3D shows the average resonance amplitude at different $d_1$ values at 0 V. The resonance depth decreases with an increase in $d_1$ and stays nearly constant for $d_1$ values larger than the critical coupling distance ($d_c$) of 2.4 mm. Indeed, we are limited with the noise level of our measurement system. Then, we measured the variation of the near-field distribution as a function of bias voltage at $d_1$=3 mm, Fig. 3E. Upon increasing the charge density on graphene electrodes, the extinction ring shrinks further and diminishes at large voltages. The transmission spectrums at 0 V and 1.2 V are demonstrated in Fig. 3F. While the size of the extinction pattern becomes smaller until a critical bias voltage ($V_c$) of 0.55 V, the resonance amplitude does not change and it is damped beyond that voltage. The critical bias voltage to start damping increases with the separation distance for $d_1 \geq d_c$. More details about the imaging near field extinction pattern of SRRs are presented in section S4.

**DISCUSSION**

To show the promises of the approach, now we would like to demonstrate the spatial tuning of metamaterials that could open new applications in transformation optics (*48*). Transformation optics is an emerging field of research aiming to conceal an object by bending or distorting electromagnetic waves using spatially varying metamaterials (*49, 50*). Ability to control the spatial variation of permittivity and permeability by electrical means would yield new vistas for the adaptive electromagnetic cloaking and camouflage. By integrating arrays of individually addressable metadevices, our approach can be used for voltage controlled adaptive transformation optics. Fig. 4 shows the schematic of the multi-pixel device. By controlling the local charge density on a pixelated surface by passive matrix addressing, we demonstrate an electrically-reconfigurable spatial-varying metadevices. We fabricated a large area metadevices with 4x4 arrays of active pixels. Each pixel contains 9 SRRs (3x3). We measured the scattering parameter at the center of the pixels and then calculated the local dielectric function. To calculate the dielectric function, we first write the S-parameters in terms of impedance and refractive index of the metasurface and calculated their roots in terms of S-parameters as given by (*51*). To overcome the branching problem for the real part of the refractive index, we used Kramers-Kronig relations as developed by (*52*). We used our unit cell size of 11mm for the effective thickness of our metamaterial structure. Fig. 4 (B and C) show the calculated real and imaginary part of the effective dielectric constant. By accumulating charges on a pixel, the local dielectric constant varies from 5 down to 0.8. Fig. 4 (D and E) show spatial maps of the real part of the dielectric constant of the pixelated device at different voltage configurations. Details for the spatially-varying metadevices are explained in section S5. We were able to reconfigure the spatial variation of the dielectric constant over the device by precisely varying the voltage applied to the rows and columns. It is necessary to point out here that the cross coupling between the pixels is due to the passive matrix addressing. The fabricated devices are on thin flexible substrates and therefore, they can be wrapped around non-planar objects for cloaking applications.

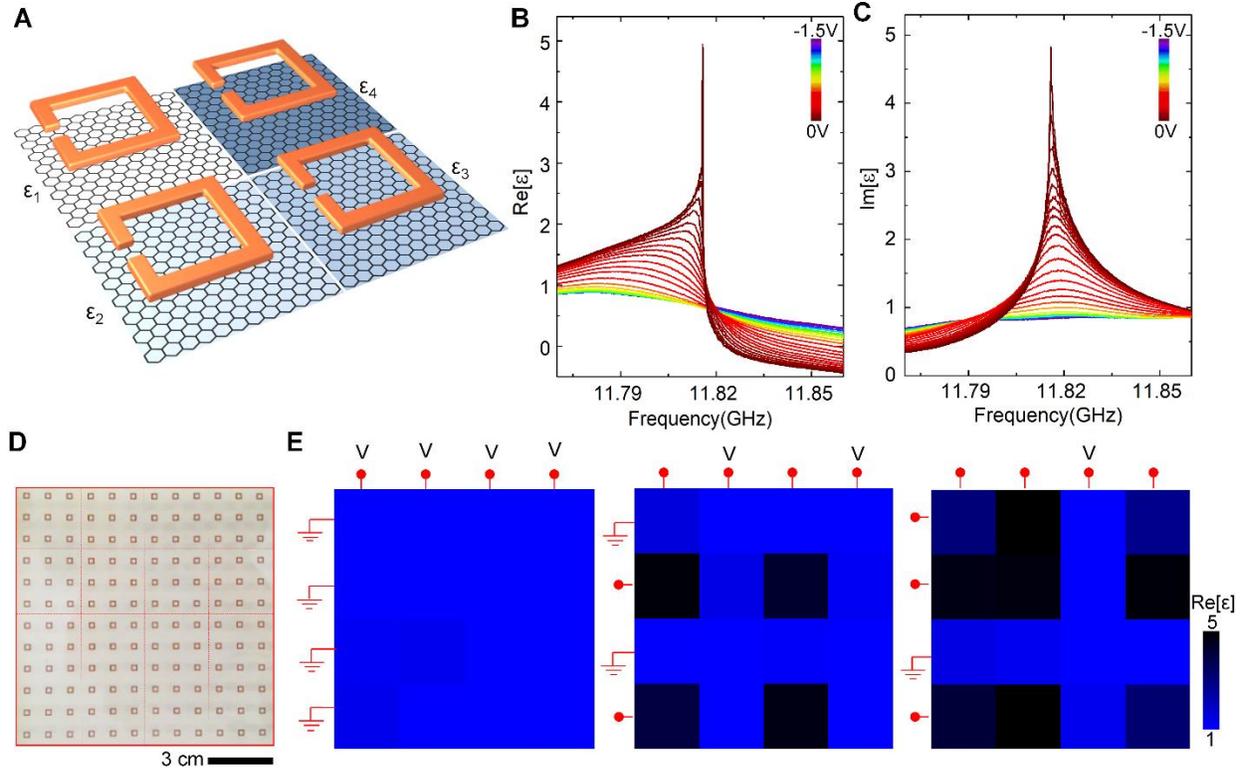

**Fig. 4. Electrically-reconfigurable spatially-varying digital metamaterials. (A)** Schematic drawing of a digital metamaterial with individually addressable pixels. Controlling the charge density on individual pixels allows us to reconfigure the local dielectric constant. **(B, C)** The spectra of the calculated real and imaginary part of the effective dielectric constant of a single pixel under different bias voltages. **(D, E)** The spatial map of the real part of the effective dielectric constant of the device at the resonance condition for different voltage configurations. The effective dielectric constant of the pixels is controlled by passive matrix addressing.

Tuning resonance frequency of a metasurface without modifying its amplitude is one of the challenges for active metadevices. To overcome this challenge, we have implemented a unique metadevice concept in order to actively control the resonance frequency of the device by electrical means. The metadevice consists of three coupled layers; first layer is SRR, and the second layer is LR arrays, and the third layer is the graphene capacitor. We used large area graphene capacitor as an electrically tunable active component of the metadevice. The unit cell of the metadevice is schematically illustrated in Fig. 5A where two SRRs coupled to a single LR at $d_2$ and both of them is coupled to a graphene capacitor at $d_1$ below the LR. It should be noted that each layer is coupled capacitively to each other and the coupling capacitance is proportional to the distance between them. Small signal circuit model of the entire metadevice with electrostatic coupling capacitances are shown in Fig. 5B. Graphene is represented with a tunable quantum capacitance ($C_q$) and sheet resistance ($R_s$). Uncoupled single layer of SRR and LR surfaces have their own resonance deeps in transmission spectrum, simulated results are shown in Fig. 5C. Coupling these layers at $d_2$=3.5 mm results in two resonance deeps; while one resonance is SRR-like the other resonance is LR-like. LR-like resonance is at higher frequencies and SRR-like resonance is at lower frequencies than the uncoupled states. Our aim here is to actively tune the SRR-like resonance frequency without changing its amplitude by selectively damping the LR-like resonance. Therefore, we placed graphene capacitor closer to the LR layer than SRR layer. Transmission spectra measured from the three layer metadevice as a function of bias voltages are shown in Fig. 5D for $d_1$=1.35 mm and $d_2$=3.50 mm. The regions with blue color show the resonance amplitude below -50 dB. Charging graphene capacitor with bias voltages damps the resonance amplitude of the LR-like resonance (Fig. 5E) while shifting the resonance frequency of the SRR-like resonance (Fig. 5F). The amplitude of the SRR-like resonance fluctuates with applied bias voltages but its overall amplitude stays below -50 dB level. We have obtained a similar active frequency tunable metadevice by coupling two uniform SRR layers (see section S6).

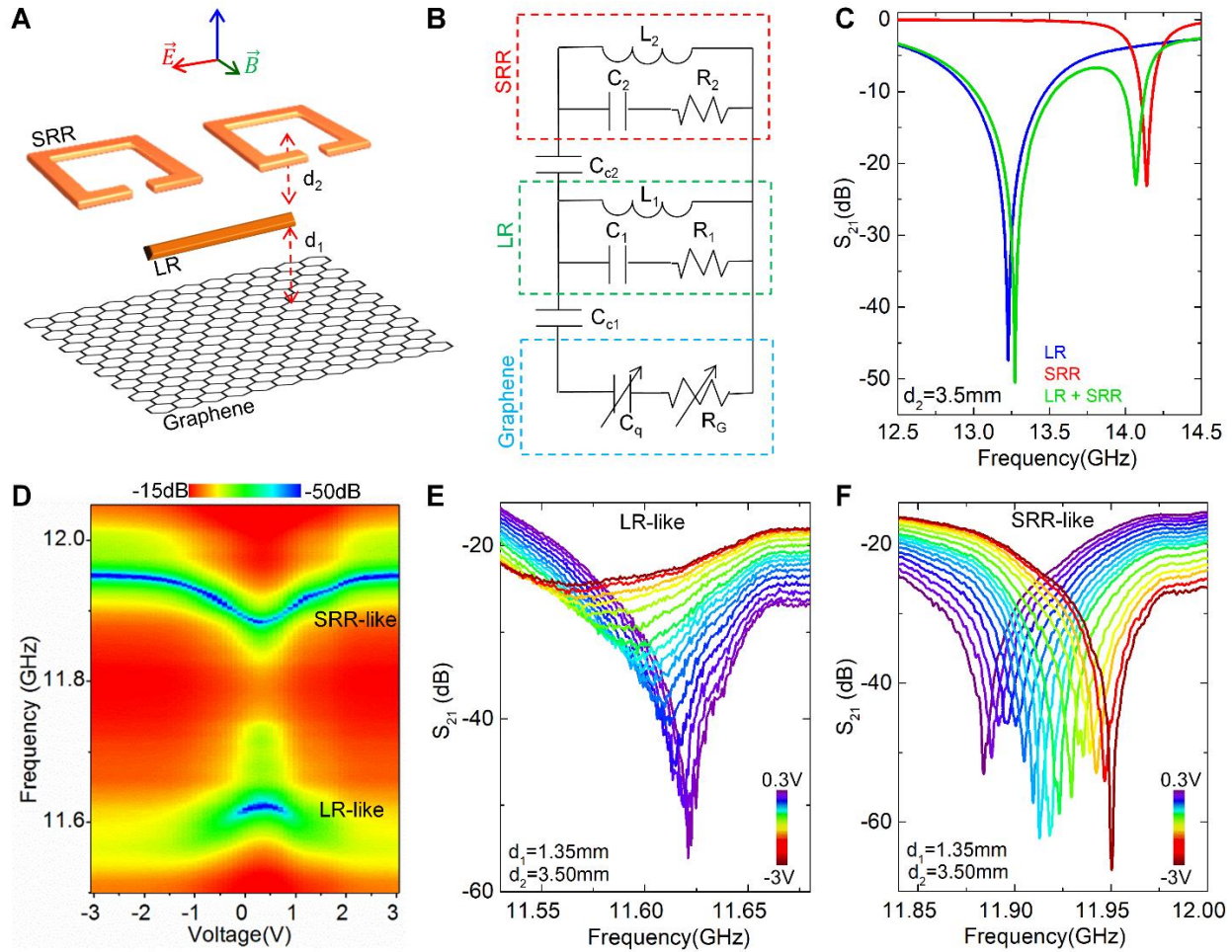

**Fig. 5. Frequency tunable active metadevices.** (**A**) Schematic drawing of the designed three layer frequency tunable metadevice structure. In a unit cell, two square ring resonators (SRRs) are coupled to a line resonator (LR) at $d_2$ and a graphene capacitor is coupled to both of them at $d_1$ below the LR layer. (**B**) Small signal circuit model to represent the three layer metadevice. (**C**) Transmission spectrum showing the resonances of coupled ($d_2$=3.5 mm) and uncoupled SRR and LR metasurfaces. (**D**) Experimentally measured variation in the transmission spectrum of the three layer metadevice by means of bias voltages applied to graphene capacitor at $d_1$=1.35 mm, $d_2$=3.50 mm. (**E**) Transmission spectrum showing the resonance damping of LR-like resonance with bias voltage. (**F**) Transmission spectrum showing the shift in the frequency of the SRR-like resonance with bias voltages without varying its amplitude.

**CONCLUSION**

The results presented here confirm a promising approach for electrically controlled active metadevices. The core idea of the approach is based on the electrical tuning of effective strength of a resonator placed close to a tunable Drude metal. Electrostatic tuning of the charge density on graphene yields broadband controllability on the response of various types of metamaterials both in near-field and far-field (section S7 and S8). These active metadevices enable efficient control of both amplitude (> 50 dB) and phase (> 90°) of electromagnetic waves. To show the promising uses of the approach, we demonstrated electrically reconfigurable spatially varying metamaterials and frequency tunable active metadevices. The operation frequency of these metadevices can be easily scaled up to the terahertz and higher frequencies. Large modulation depth, simple device architecture, and mechanical flexibility are the key attributes of the graphene-enabled active metadevices. We anticipate that the presented approach could lead to new applications ranging from electrically switchable cloaking devices to adaptive camouflage systems in microwave and terahertz frequencies.

**MATERIALS AND METHODS**

**Fabrication of graphene capacitors**

We synthesized large area (7x14 cm$^2$) graphene using a chemical vapor deposition system on ultra-smooth copper foils (Mitsui mining and smelting company, LTD, B1-SBS) under 100 sccm H$_2$ and 10 sccm CH$_4$ flow for 1 min. at 1035°C. After the growth, we transferred the graphene on a flexible 75 μm thick PVC substrate using hot lamination technique. After etching the Cu in 5 mM aqueous solution of FeCl$_3$, we used two layers of graphene on PVC to fabricate a large area graphene supercapacitor. A paper tissue with a thickness of 100 μm was used as a spacer between two graphene electrodes. We first put the tissue on top of one graphene electrode and then we put ionic liquid (1-Butyl-3methlimidazolium hexafluorophosphate). We placed the top graphene electrode, and let the ionic liquid disperse between the two graphene electrodes.

**Fabrication of metamaterials**

To fabricate metamaterial structures, we laminated 75 µm thick PVC laminating film on 10 µm thick copper foils. We printed the metamaterial structure on the copper foils using HP color laser jet printer (CP2020). We etched the copper in a nitric acid solution. The printed toner function as an etch mask on copper. With this technique we can define metadevices having a feature size as small as 80 µm.

**Microwave measurements**

We used a Keysight-E5063A network analyzer for the microwave characterization of our device. For near-field measurements; we connected a horn antenna to one channel and a 1.5 cm long monopole antenna to the other channel of the network analyzer to measure S-parameters in the 7-15 GHz range. The active metadevice was placed at a specific distance above the monopole antenna and the horn antenna was used to propagate microwaves in free space. The device is placed on a computer controlled motorized stage with 1 um spatial resolution. Both the amplitudes and phases of the received signals were recorded for each location. The monopole antenna was aligned along the electric field of the horn antenna. We connected the Keithley 2400 source measure unit to apply voltage bias. We tuned the voltage manually and recorded the scattering parameters. For far field measurements we used a horn antenna instead of a monopole antenna and placed our devices between two horn antennas to measure far field S-parameters.

**SUPPLEMENTARY MATERIALS**

section S1. Fabrication and characterization of graphene super-capacitors

section S2. Microwave performance of active meta-devices in near-field

section S3. Electromagnetic modelling and simulation of active Metadevices

section S4. Imaging near-field extinction pattern of a square SRR

section S5. Electrically-reconfigurable spatially-varying digital metamaterials section S6. Frequency tunable active metadevices section S7. Performance of active metadevices in far-field section S8. Microwave performance of various metadevices in near-field fig. S1. Picture of growth, transfer and etching processes.

fig. S2. Characterization of single layer graphene.

fig. S29. Imaginary part of patterned dielectric constant on a large area metadevice.

fig. S30. Voltage response of transmission ($S_{21}$) and retrieved dielectric constants in pixelated metadevices.

fig. S31. SRR arrays coupled to line resonator arrays.

fig. S32. Finding optimum $d_1$ to achieve maximum shift in resonance frequency of SRRs at $d_2$=3.5 mm.

fig. S33. Transmission spectrums showing the shift in resonance frequency of SRR arrays while damping the resonance amplitude of line resonators.

fig. S34. Coupling two identical SRR arrays.

fig. S35. Tuning resonance frequency of SRR arrays coupled to its twin.

fig. S36. Active tuning of magnetic resonance.

fig. S37. Active tuning of electric resonance excited by a horizontally travelling electromagnetic waves on split ring resonator arrays by graphene supercapacitor.

fig. S38. Electrically switchable metadevices made of a dense SRR arrays measured in far-field.

fig. S39. Switching transmission amplitude and phase of various type metamaterials coupled to graphene supercapacitors by sweeping bias voltage from 0 V to -1.5 V in near-filed.

fig. S40. Switching transmission amplitude and phase of various type metamaterials coupled to graphene supercapacitors by sweeping bias voltage from 0 V to -1.5 V measured in near-filed.

fig. S41. Schematic drawings of metamaterials fabricated throughout this work with their dimensions.

## REFERANCES AND NOTES

**Acknowledgments**

**Funding:** This work was supported by the European Research Council (ERC) Consolidator Grant No. ERC-682723 SmartGraphene and by the Scientific and Technological Research Council of Turkey (TUBITAK) Grant No. 114F052. O.B. and N.K. acknowledge the fellowship provided by TUBITAK-BIDEB 2211 and 2215. **Author contributions:** O.B. and C.K. conceived the idea and planned the experiments. O.B. fabricated the samples, performed the experiments and made the simulations. O.B. and C.K. analyzed the data and wrote the manuscript. N.K., S.B. and E.O.P. helped for the synthesis of graphene, fabrication of the devices and microwave measurements. E.K., S.C., H.C. and E.O. helped for the simulations and interpretations of the experimental results. All authors discussed the results and contributed to the scientific interpretation as well as to the writing of the manuscript. **Competing interests:** The authors declare that they have no competing interests. **Data and materials availability:** All data needed to evaluate the conclusions in the paper are present in the paper and/or the Supplementary Materials. Additional data related to this paper may be requested from the authors.